\pdfoutput=1
\documentclass[11pt]{article}

\usepackage[]{acl}

\usepackage{times}
\usepackage{latexsym}
\usepackage{amsfonts}
\usepackage{verbatim}
\usepackage{amsmath}

\providecommand{\mailto}[1]{\href{mailto:#1}{\texttt{\textls[-30]{#1}}}}    % provide email command

\usepackage[ruled,vlined]{algorithm2e}
\RestyleAlgo{ruled}

\usepackage[T1]{fontenc}
\usepackage[utf8]{inputenc}

\usepackage{microtype}

\usepackage[disable]{easy-todo} %disable
\usepackage{booktabs}
\usepackage{subcaption}
\usepackage{csquotes}
\usepackage{amsmath,mathtools,bm}
\usepackage[nameinlink,capitalise]{cleveref}

\usepackage{xspace}
\usepackage{glossaries}
\glsdisablehyper
\newacronym{DP}{DP}{differential privacy}

\newcommand*{\Dlp}{\Gls{DP}\xspace}
\newcommand*{\dlp}{\gls{DP}\xspace}
\newcommand*{\diffpriv}{\acrlong{DP}\xspace}

\newcommand*{\dyp}{differentially private\xspace}

\newcommand*{\mech}{mechanism\xspace}
\newcommand*{\randmech}{random mechanism\xspace}

\newacronym[longplural={Bags-of-Words}]{BoW}{BoW}{Bag-of-Words}
\newcommand{\bow}{\gls{BoW}\xspace}

\newtheorem{definition}{Definition}

\newcommand*{\eps}{\ensuremath{\epsilon}\xspace}
\newcommand*{\del}{\ensuremath{\delta}\xspace}
\DeclareMathOperator{\adj}{\sim}

\DeclarePairedDelimiter{\abs}{\lvert}{\rvert}
\DeclarePairedDelimiter{\norm}{\lVert}{\rVert}
\DeclarePairedDelimiter{\parens}{\lparen}{\rparen}
\DeclarePairedDelimiter{\brackets}{\lbrack}{\rbrack}

\newcommand{\pr}[2][]{\Pr\brackets[#1]{#2}}

\newcommand{\dist}{d}

\newcommand{\dEdit}{\ensuremath{\dist_{\pm1}}}

\providecommand{\D}{\mathcal{D}}
\providecommand{\R}{\mathcal{R}}
\providecommand{\X}{\mathcal{X}}
\providecommand{\Y}{\mathcal{Y}}

\providecommand{\numberset}[1]{\ensuremath{\mathbb{#1}}}
\providecommand{\Reals}{\numberset{R}} % reals
\newcommand{\vv}[1]{\bm{#1}}		% vectors
\newcommand{\vx}{\vv x}
\newcommand{\vy}{\vv y}
\newcommand{\vz}{\vv z}

\newcommand{\Mech}[1]{\mathcal{#1}}
\providecommand{\M}{\Mech{M}} % any (random) mechanism

\DeclareMathOperator{\supp}{supp}	% support

\DeclareMathOperator{\softmax}{\ensuremath{\mathrm{softmax}}}

\newcommand{\V}{\ensuremath{\mathcal{V}}}

\title{The Limits of Word Level Differential Privacy}

\author{Justus Mattern \\ RWTH Aachen University \\  \mailto{justus.mattern@rwth-aachen.de}
  \And Benjamin Weggenmann \\ SAP Security Research \\ \mailto{benjamin.weggenmann@sap.com}
  \AND Florian Kerschbaum \\ University of Waterloo \\ \mailto{florian.kerschbaum@uwaterloo.ca}
} % email addresses are too long to fit all authors on one line :/
\begin{document}
\maketitle
\begin{abstract}
As the issues of privacy and trust are receiving increasing attention within the research community, various attempts have been made to anonymize textual data. A significant subset of these approaches incorporate differentially private mechanisms to perturb word embeddings, thus replacing individual words in a sentence. While these methods represent very important contributions, have various advantages over other techniques and do show anonymization capabilities,
they have several shortcomings. In this paper, we investigate these weaknesses and demonstrate significant mathematical constraints diminishing the theoretical privacy guarantee
as well as major practical shortcomings with regard to the protection against deanonymization attacks, the preservation of content of the original sentences as well as the quality of the language output. Finally, we propose a new method for text anonymization based on transformer based language models fine-tuned for paraphrasing that circumvents most of the identified weaknesses and also offers a formal privacy guarantee. We evaluate the performance of our method via thorough experimentation and demonstrate superior performance over the discussed mechanisms.

\end{abstract}

\section{Introduction}
\label{sec:intro}

Computational authorship attribution approaches ranging from rule-based methods measuring character-level $n$-gram frequencies \citep{kevselj2003n} to models incorporating deep learning \citep{shrestha2017convolutional} make it possible to identify the authors of a given text. While these technologies enable valuable applications such as supporting historians in their research,  they can potentially be exploited by attackers to identify the originators of sensitive data and thus diminish the privacy of individuals. To protect the anonymity of users whose data is being shared online and used by companies and researchers, methods that anonymize the writer of given texts are necessary and of interest within the research community and a variety of industries, specifically those handling personal information such as healthcare or financial services.

Previous work in the field of authorship obfuscation mainly focuses on two different tasks, namely learning anonymous textual vector representations for downstream tasks \citep{coavoux2018privacy, weggenmann2018syntf, fernandes2019generalised, mosallanezhad2019deep, beigi2019i} and the development of mechanisms that transform the input sentence to remove properties revealing the author and thus output human-readable text.
Works within the second category \citep{feyisetan2019leveraging,feyisetan2020privacy,DBLP:journals/corr/abs-2010-11947, bo-etal-2021-er} typically follow a common \emph{word level} framework which is characterized by the differentially private individual perturbation of word embeddings and the subsequent sampling of new words that are close to the perturbed vectors in the embedding space.
Also, the majority of recent work proposing new methods for authorship obfuscation deals with the optimization and calibration of noise sampling mechanisms \citep{xu2020differentially} or the definition of new distributions to sample noise from \citep{feyisetan2019leveraging} as opposed to the development of entirely new methods. 
%We argue that the word level framework faster progress would be enabled by reevaluating the framework of word level differential privacy and the incorporation of completely new architectures.

In this paper, we thoroughly investigate the capabilities of word level anonymization from the theoretical perspective of \dlp (\cref{sec:dpconstraints}), in terms of the language quality of its output (\cref{sec:languageconstraints}) as well as from a utilitarian perspective considering the ability to protect the privacy of people whose data is being used. Specifically, we extend the experimentation in papers proposing the discussed methods by testing their capability to mitigate deanonymization attacks using state-of-the-art methods on the widely used IMDb movie review and Yelp business review datasets (\cref{sec:eval}).
We find that the technical constraints applied to fulfill \dlp in the local model cause strong limitations,
and, more importantly, observe that, despite the formal guarantees, such methods offer little protection against advanced deanonymization attacks.
For this reason, we advocate for approaches granting more flexibility to the text generation process and, motivated by experiments showing that human rewritings of texts gathered through crowdsourcing successfully anonymize the original authors \citep{mishari-crowdsourcing}, propose an anonymization approach based on paraphrasing (\cref{sec:paraphrasing}) that maintains the advantages and the theoretical privacy guarantee of the discussed methods, evades most of the identified drawbacks and outperforms word level mechanisms in our experiments.
%\todo{include forward references (cref) to respective sections}

\section{Background}
\label{sec:bg}

%\begin{itemize}
%    \item Describe differential privacy
%    \item Describe framework of word level privacy (general %algorithm)
%    \item Papers for word level DP:
%    \begin{enumerate}
%    \item Laplace Noise on euclidian Glove embeddings: %%\citep{feyisetan2020privacy}
%    \item hyperbolic noise on Poincaré embeddings: \citep{feyisetan2019leveraging}
%    \item Elliptic noise with scale regularized by Mahalanobis norm on euclidian embeddings (GloVe): \citep{DBLP:journals/corr/abs-2010-11947}
%    \end{enumerate}
%\end{itemize}

The majority of proposed text anonymization methods rely on a common framework that applies \dlp on a per-word level by perturbing individual word embeddings \citep{feyisetan2019leveraging, feyisetan2020privacy, DBLP:journals/corr/abs-2010-11947, xu2020differentially, xu2021utilitarian}. In this section, we introduce the concept of \dlp and give an overview of the commonly used word level framework.

\subsection{Differential Privacy}
\label{sec:bg:dp}

\Dlp has been introduced by \citet{dwork2006calibrating} under the name \emph{\eps-indistinguishability}.
Its goal is to give semantic privacy by quantifying the risk of an individual
that results from participation in data collection.
In the original, \emph{central model}, we assume the collected data is stored
in a central database with one record per participant.
If we consider \emph{adjacent} databases that differ by at most one record (pertaining to one individual),
%(i.e., one individual's data),
a \dyp query on both databases should yield matching results with similar probabilities,
i.e., answers that are probabilistically \emph{indistinguishable}.
This is achieved via \emph{random mechanisms} on the universe of datasets $\D$
that return noisy query results, thus masking the impact of each individual.

% In terms of Bayesian notation,
%A random mechanism $\M(\rx)$ with input $\rx$ is characterized
%by a conditional distribution $p_\M(\rz \mid \rx)$
%where $\rz$ is a random vector representing its output;
%we then say $\M$ is the mechanism \emph{induced by $p_\M(\rz \mid \rx)$}.
%We \emph{run} a \randmech $\M$ on a given input $\rx$
%by sampling a realization $\rz$ from $p_\M(\rz \mid \rx)$.
%Depending on the context, we overload the notation so that $\M(\rx)$
%can also be interpreted as its underlying distribution.

\begin{definition}[\eps-\dlp]
    \label{def:diffpriv}
    Let $\eps>0$ be a privacy parameter.
    A \randmech $\M:\D\to\R$ fulfills \emph{\eps-\dlp} %is \emph{$\epsilon$\-/\dyp}
    if for all adjacent databases $D,D'\in\D$, and all sets of possible outputs $R \subset \supp \M$,
    \[
        \Pr\brackets{\M(D)\in R} \leq e^\epsilon \cdot \Pr\brackets{\M(D') \in R}.
    \]
\end{definition}

To make a query function $f:\D\to\R$ \dyp, %in the central model, 
noise is calibrated to the query's \emph{sensitivity},
i.e. its maximal change over all pairs of adjacent datasets $D \sim D'\in\D$.
% In case of isotropic Gaussian mechanisms,
For instance, the L2-sensitivity as used for the Planar Laplace mechanism
\cite{chatzikokolakis_2013_broadening,andres2013geo,koufogiannis2015optimality}
is % and Gaussian mechanisms
\[
    \Delta_2 f := \max_{D\sim D'} \norm{f(D)-f(D')}_2.
\]

In the \emph{local model} \citep{duchi2013local},
noise is added locally at the data source, before the data is collected
and stored in a central database.
A basic example is randomized response \citep{warner1965randomized},
where each survey participant either provides a truthful or a random answer
depending on the flip of an (unbiased) coin.
The local model makes the strong assumption that any two inputs are considered adjacent,
which often makes it difficult to achieve a satisfying privacy-utility trade-off.

% ---------------------------------------------------------------------------------------------
\subsubsection{Generalization with metrics}
% ---------------------------------------------------------------------------------------------
\label{sec:bg:dp:metrics}

% A limitation with differentialprivacy is that the indistinguishability is obtained
% on a per-record level regardless of the actual values of the differing records.
A limitation with \dlp is that the indistinguishability is achieved
between two inputs on a per-record level regardless of their actual values.
This can be especially problematic in the local model,
where each user might just submit one single record,
in which case a \dlp mechanism with small privacy parameter \eps
would enforce each submitted record to be indistinguishable from any other,
thus rendering the collected data essentially useless.
\citet{chatzikokolakis_2013_broadening} argue that in some scenarios,
the (in)distinguishability between two databases as enforced by a privacy mechanism
should depend on the values themselves instead of the number of differing records.
% We might require, for instance, databases in which the value of an
% individual is only slightly modified to be highly indistinguishable, thus protecting the
% accuracy by which an analyst can infer an individual’s value.
They hence propose a generalized notion of \emph{privacy on metric spaces}
% that also covers domains other than databases
% which might lack a natural notion of adjacency, but
where a mechanism run on \emph{nearby} elements %$\vx,\vx'$ 
results in \emph{similar} output probabilities:
\begin{definition}[Metric privacy]
	\label{def:metric-privacy}
  Let $\eps>0$ be a privacy parameter.
  On a metric space $(\X,\dist)$, a mechanism $\M$ satisfies \emph{$\eps\dist$-privacy}
  if for all $\vx,\vx'\in\X$ and all $R\subset\supp\M$,
	\[
		\Pr\brackets{\M(\vx) \in R} \leq e^{\eps \cdot \dist(\vx,\vx')} \cdot \Pr\brackets{\M(\vx') \in R}.
	\]
\end{definition}
% \todo{Introduce and explain relation between protection radius and level!}
In other words, the indistinguishability level of two points $\vx,\vx'$
is bounded by $\eps$ times their distance.
% amounts to $\eps \dist(\vx,\vx')$, i.e. it depends on \eps and their distance.
% For an $\eps\dist$-private mechanism $\M$

% % Not required for now.
% \citet{andres2013geo} provide another interpretation:
% If we consider an arbitrary but fixed distance $r>0$,
% any two points with $\dX(\vx,\vx')\leq r$
% achieve a level of indistinguishability at most $\ell:=\eps r$.
% Thus, an $\eps\dX$-private mechanism $\M$ achieves
% a \emph{privacy level} $\ell=\eps r$ within a \emph{protection radius} $r$.
% % (cf.~\cite{andres2013geo}).

Note that we recover the original notion of central \dlp on the space of databases $\X=\mathcal{D}$
if we use the \emph{record-level edit distance} $\dEdit$, as datasets $\vx,\vx'\in\mathcal{D}$ 
differ by at most one record if and only if $\dEdit(\vx,\vx') \leq 1$.
Similarly, the local model is obtained for $\dist(\vx,\vx') \equiv 1$.
This motivates the following broader and formal definition of adjacency:
%Two databases $x_1,x_2\in\X$ that differ in exactly one record are called \emph{adjacent};
%these are precisely those whose Hamming distance $\dHamming(x_1,x_2)$ is $1$.
%One important notion in this context is that of \textit{adjacency} between two inputs:
\begin{definition}%[Adjacency]
  \label{def:adjacency}
  In a metric space $(\X,\dist)$, we call two inputs $\vx,\vx'\in\X$
  \emph{adjacent (with respect to $\dist$)} if $\dist(\vx,\vx')\leq1$.
  We write this as $\vx \adj_{\dist} \vx'$ (or $\vx \adj \vx'$ if $\dist$ is understood from the context).
\end{definition}

\subsection{Word perturbations for anonymization}
\label{sec:bg:word level}

The methods investigated in this paper apply word embedding perturbation mechanisms to change individual words in a sentence, following $\eps\dist$-privacy with a distance metric defined for sentences $\vx,\vx'$.
In essence, the common word level framework works as follows:
Given an input sentence \(\vx = (x_1, x_2, ..., x_n)\), each token \(x_i\) is mapped to an $n$-dimensional pretrained word embedding \(\phi(x_i)\). Subsequently, an $n$-dimensional noise vector \(\eta\) is sampled from a multivariate probability distribution \(p_\epsilon(\eta)\) and added to the word embedding to obtain a noisy vector \(\hat{\phi}_i\). The current word \(x_i\) then gets replaced by a word \(x'_i\) whose embedding \(\phi(x'_i)\) is close to the noisy embedding \(\hat{\phi}_i\).
%\todo{For Poincaré, shouldn't we use some hyperbolic metric instead of Euclidean? In that case, maybe just stating that the "word should be close (in whatever embedding space)" is enough.}
%in terms of Euclidean distance.
Given a distance metric \(d\), commonly \(d(\vx,\vx') = \sum_{i=1}^{n} \norm{\phi(x_i)-\phi({x'_i})}\) for sentences \(\vx, \vx'\) of the same length, the mechanism fulfills $\eps\dist$-privacy. The general mechanism is outlined in \cref{alg:cap} and the proofs are outlined in the referenced papers.

\begin{algorithm}
  \microtypesetup{protrusion=false} % fix missing space after \KwInOut keywords
  \small
  \caption{Word level DP framework}\label{alg:cap}
  \label{alg:wordleveldp}
  \SetKwInOut{KwIn}{Input}
  \SetKwInOut{KwOut}{Output}
  \KwIn{Text \(\vx = (x_1, x_2, \ldots, x_n)\), parameter \(\epsilon\)}
  \KwOut{Anonymized text \(\vx' = (x'_1, x'_2, ..., x'_n)\)}

  %\For{\(i \in \{1, 2, ..., l\}\)}{
   % \State Compute embedding \phi_i = \phi(x_i)\\
%    \State Sample noise \(z \sim p_\epsilon(z)\)\\
%    \State Compute perturbed embedding \hat{\phi}_i = \phi_i + z\\
%    \State Obtain close word \(\hat{x}_i\) with distance ||\phi(\hat{x}_i) - \hat{\phi}_i||\\
%    \State Insert \(\hat{x}_i\) {for} \(x_i\)}
  \For{\(i \in \{1, 2, \ldots, n\}\)}{
   Compute embedding $\phi_i = \phi(x_i)$\\
   Sample noise $\eta \sim p_\epsilon(\eta)$\\
   Compute perturbed embedding $\hat{\phi}_i = \phi_i + \eta$\\
   Find near word \(x'_i\) within embedding space\\
   Insert $x'_i$ for $x_i$ in the output\\
  }
\end{algorithm}

\section{Limitations of word level privacy}
\label{sec:limitations}

% This section investigates the framework of word level differential privacy from a theoretical and practical standpoint. We examine theoretical limits from two perspectives. First, we identify and map out significant mathematical constraints and assumptions that have to be made for word level mechanisms to fulfill differential privacy, thus diminishing the actual privacy guarantee provided by DP. Secondly, we investigate the language production capabilities by the anonymization mechanisms studied in this paper and identify strong weaknesses with respect to language quality, anonymization capabilities and content as well as utility preservation.
% Beside these theoretical contributions, we test named anonymization mechanisms experimentally and measure their usefulness for real systems and protection capabilities against an adversary.
% \todo{Revert to previous intro text?}

% \Dlp (cf.\ \cref{sec:bg:dp}) is a notion of privacy based on \emph{randomness},
% i.e., any non-trivial \dlp mechanism must be non-deterministic.
%
\dlp mechanisms operating on a word-by-word basis follow a comparably simpler and more straightforward algorithmic approach than deep learning models for text anonymization. This has many advantages such as lower computational expense as well as the mechanism's independence of the target dataset and domain: Most deep learning based approaches need to be trained for each dataset and set of authors individually as they require author labels to construct adversarial training objectives \citep{shetty2018a4nt, xu2019privacy}. In contrast, the approaches discussed in this paper are dataset-independent and can thus be deployed immediately without a need for further training for new authors and datasets.

The simple methodology does however have its shortcomings as well. In this section, we examine these weaknesses from a theoretical standpoint taking into account both \dlp properties and properties of the language output before assessing their effects experimentally in \cref{sec:eval}.

\subsection{DP related constraints}
\label{sec:dpconstraints}

% Overall: Independent word-by-word processing

% Output constraints
% - unordered, non-readable output (e.g., BoW model)
% - fixed length sequences
% - (non-DP cause) incoherent output (requires large eps to keep sentence intact -- see linguistic limits)

% Privacy budget
% - grows linearly in length of the text

We consider a mechanism $\M$ that operates on a text $\vx = \parens{x_1,\ldots,x_n}$
on a word-by-word basis, i.e., $\M(\vx) = \parens{\M(x_1),\ldots,\M(x_n)}$.
% by slightly overloading the notation.

\paragraph{Length constraints}
A word level mechanism $\M$ will produce an output that has the same length as its input.
However, typical texts and sentences come in varying lengths,
say $\vx = \parens{x_1,\ldots,x_n}$ and $\vx' = \parens{x'_1,\ldots,x'_m}$ with $n\neq m$.
Now if we consider an outcome set $Z_n$ consisting of all length-$n$ sequences (including $\vx$),
we obtain
\[
  1 = \pr{\M(x)\in Z_n} \not\leq e^\eps \pr{\M(x')\in Z_n}=0.
\]
This contradicts the definition of pure \dlp and in case of approximate \dlp (cf.\ \cref{def:diffpriv})
would require $\del=1$ which is clearly not negligible.

To comply with these strong \dlp requirements, word level \dlp mechanisms such as
\citet{feyisetan2019leveraging,feyisetan2020privacy} commonly %cope with this %,bo-etal-2021-er
simply limit the privacy guarantee to cover only sequences $Z_n$ of a fixed length $n$,
i.e., no formal guarantee among sentences of different lengths is provided.
Consequently, the output is also fixed to length $n$,
which affects the language capabilities of such mechanisms
% This is used e.g. by \citet{feyisetan2019leveraging,feyisetan2020privacy,bo-etal-2021-er},
and severely limits the scope and expressiveness of the resulting sentences,
particularly for human readers.
%\todo{May be sufficient to mention this here only, or should we mention this (rather / again) in the "linguistic limitations" subsection? Yes, definitely sufficient -Justus}
% \todo{\textcolor{blue}{Limiting the output is inherent to the mechansim's defintion, so isn't the actual coping strategy limiting the definition of the space of sentences in the DP formulation? This is in my opinion more drastic as this means the privacy guarantee (indistinguishability) only exists for sentences with one length n, which is a weird assumption for real world scenarios. The limitation of fixed length output is of course relevant, but the expressiveness problem is then a part of language constraints}}.

\paragraph{Linear growth of privacy budget} % -- poor privacy--utility tradeoff
% \paragraph{Undesirable impact of the local model}
For an \eps-\dlp mechanism $\M$, its output probabilities given two adjacent inputs
have to be bounded by $\exp(\eps)$.
% Consider an \eps-\DP mechanism $\M$ that
Suppose $\M$ processes each word $x_i$ of a text $\vx = \parens{x_1,\ldots,x_n}$ independently,
using a fixed-length output strategy as described in the preceding section,
with a given output $\vz=\parens{z_1,\ldots,z_n}$.
Then $ \pr{\M(\vx) = \vz} = \prod_{i=1}^n p_i $ where $p_i := \pr{\M(x_i)=z_i}$.
Similarly, a second text $\vx'$ has output probabilities $p_i' = \pr{\M(x_i)=z_i}$,
so we have $p_i \leq e^\eps p_i'$, and hence
% where $p_i := \pr{\M(x_i)=z_i}$. % and $p'_i := \pr{\M(x'_i)=z_i}$.
\begin{align*}
  \pr{\M(\vx) = \vz}
  &= \prod_{i=1}^n p_i
  \leq \prod_{i=1}^n e^\eps p_i' \\
  %&= e^{n\eps} \prod_{i=1}^n \pr{\M(\vx') = \vz}.
    &= e^{n\eps} \pr{\M(\vx') = \vz}.
\end{align*}
Therefore, the total privacy budget required by $\M$ to privatize the entire sequence
is bounded by $n\eps$ and thus may grow linearly with its length.

Metric privacy \emph{hides} this effect \emph{in} the metric,
since deviations in the mechanism's output probabilities are bounded
by $\exp\parens{\eps\dist(\vx,\bm{\vx'})}$.
By choosing a metric $\dist$ that grows larger as the length of sentences increases,
strong deviations can now be captured by the metric $\dist$,
so the privacy budget \eps as its co-factor \emph{appears} smaller.
For instance, \citet{feyisetan2020privacy} use a metric
$\dist(\vx,\bm{\vx'}) = \sum\norm{\phi(x_i) - \phi( x'_i)}$
for strings based on embeddings $\phi$,
which results in more summands and thus larger distances for longer strings, but not necessarily larger distances for different writing styles: Consider the following sentence pairs \((\vx,\vx')\) and \((\vy,\vy')\) written by two authors each:
\begin{align*}
  \vx  &= \text{\enquote{Today I feel great}}\\
  \vx' &= \text{\enquote{I feel great today}}\\
  \vy  &= \text{\enquote{Today I feel great and will get a coffee}}\\
  \vy' &= \text{\enquote{I feel great and will get a coffee today}}
  % \vx = (\text{'Today'},\text{'I'}, \text{'feel'}, \text{'great'})\\
  % \vx ' = (\text{'I'}, \text{'feel'}, \text{'great'},\text{'today'})\\
  % \vy = (\text{'Today'},\text{'I'}, \text{'feel'}, \text{'great'},\text{'and'}, \text{'will'},\\ \text{'get'}, \text{'an'}, \text{'ice'}, \text{'cream'})\\
  % \vy ' = (\text{'I'}, \text{'feel'}, \text{'great'},\text{'and'}, \text{'will'}, \text{'get'},\\ \text{'an'}, \text{'ice'}, \text{'cream'}, \text{'today'})
\end{align*}
Given a non-degenerate metric $\dist$, we have both $\dist(\vx,\vx'),\dist(\vy,\vy')>0$
since the sentences are syntactically different.
One could infer that the author of the first sentence within both pairs tends to put
expressions of time in the beginning whereas the other author places them at the end,
but beyond that, there are arguably no differences in terms of writing style or author-revealing information
one could deduce from both sentence pairs.
Yet, we will likely have \(d(\vx,\vx') < d(\vy,\vy')\) due to the induced growth of the distance for longer sentences.
Hence, while the distance metric does reflect differences between sentences in a somewhat meaningful way,
it is prone to absorb the actual privacy loss even if the writing style is almost unchanged,
thus leading to values of \eps that are \emph{perceived} as small.

%\(\vx =\)"Today I feel great!" and \(\vx' =\) "I feel great today!" as well as \(\vy =\)"Today I feel great and want to buy an ice cream!" and \(\vy' =\)"I feel great and want to buy an ice cream today!". For both sentence pairs, $\dist(\vx,\bm{\vx'})$}
% For normal \dlp, this would otherwise have to be captured alone by the privacy loss parameter \eps.

% \paragraph{Curse of the local model}
\paragraph{Shortcomings of the local model}

% 1 how text is used?
In many likely scenarios for authorship obfuscation methods,
the intention is to share obfuscated texts with other,
benign entities for further processing.
For a \dlp mechanism, this essentially corresponds to the local model
where it transforms each text individually to an obfuscated output.
The assumption~then is that the obfuscation allows only privacy-insensitive processing
so that subsequent results and inferences do not harm the privacy of the texts' authors.

Note that the \dlp guarantee in the local model differs substantially
from what is expressed by the definition in the original central model: 
A central \dlp \mech would aggregate the texts (records)
from multiple individuals into a single result.
By the definition of adjacency, central \dlp hides the impact of each individual's contribution 
in the result by making it probabilistically indistinguishable (as determined by \eps)
whether an outcome was obtained with or without an individual's data.
% and the privacy guarantee is about hiding the impact of one individual in that result.
In contrast, for local \dlp, any two inputs are considered adjacent,
% each outcome depends precisely on one input (e.g., one text), namely of one individual.
so by definition, it needs to be indistinguishable whether an output
was produced by one input or another.
This strong condition makes it thus questionable if such data is still useful for an analyst.

% 4 implication on usage
Due to the nature of local \dlp, it typically introduces large amounts of noise
and requires large amounts of data to still get meaningful results
\cite{wood2020designingaccess}.
A workaround often used in practice when only limited data is available
is to use a larger privacy budget \eps than one would normally consider
sufficiently privacy-preserving in the central model
\cite{qin2016heavy,damien_desfontaines_local_2021}.
% \scriptsize \textcolor{blue}{Is this discussed in security/privacy literature?
% If so, citations could strengthen the point}\normalsize
While this does permit the obfuscated data to remain useful to a benign analyst,
it may also be useful to an attacker to infer privacy-sensitive information,
as the formal guarantee of local \dlp does not specifically prevent
such undesired or malicious inferences, especially when \eps is large.
% Note that both the benign analyst and the adversary are given the same resulting texts
% that received the same (amount of) perturbations.

% What about metric privacy?
To alleviate the strictness and implications imposed by the local model,
some approaches refer to metric privacy \cite{chatzikokolakis_2013_broadening}
as generalization of the original definition.
Metric privacy (cf.\ \cref{def:metric-privacy})
brings about a change in the definition how the privacy loss \eps is interpreted
in relation to the introduced metric and normally leads to seemingly smaller \eps values;
however, changing to metric privacy by itself does not imply any change % or improvements
to the inner workings of the mechanism.
We hence argue that it is less an improvement, but more a relaxation of the privacy guarantee
that still shares the same fundamental criticism of local \dlp,
e.g., our observation at the start of this section
where the metric grows with the length of the text
and thus hides the linear growth of the privacy budget.

% \paragraph{Incoherence due to word-by-word perturbations}
% \todo{Already explained in next section :).}

\subsection{Language constraints}
\label{sec:languageconstraints}

Aside from weaknesses concerning the privacy guarantee of \Dlp, mechanisms operating on a per-word level pose two significant shortcomings in terms of their language generation capability. First, smaller privacy budgets resulting in stronger noise added to the original data tend to cause a high amount of grammatical errors. Secondly, the lack of syntactic changes to the original sentences caused by the nature of such mechanisms considerably limits the linguistic variety and thus opportunities to deceive an adversary and provide anonymity for the authors of the texts.

\paragraph{Grammatical errors increase as privacy budget shrinks}
Word level mechanisms perturb every token \(x_i\) in a sentence independently of the rest of the text as opposed to common autoregressive sequence-to-sequence models where
\(p(x_i) = p(x_i \mid x_{i-1}, \ldots ,x_1)\).
This makes it difficult to maintain consistency and renders them unable to rectify grammatical errors induced by replacing a word with one of a different word type, e.g., a noun with an adjective.
To estimate the effect of this, we approximate the likelihood of word type exchanges for various \eps values: Using the WordNet database\footnote{Terms of use and license information: \cref{sec:dataterms}} \citep{miller1995wordnet,fellbaum2010wordnet}, we assign words from the GloVe vocabulary \citep{pennington-etal-2014-glove} one or multiple of the word type labels \emph{adjective}, \emph{adverb}, \emph{noun} and \emph{verb}. Subsequently, we apply the word perturbation mechanism proposed by \citet{feyisetan2020privacy} on a randomly selected set of 1,000 tokens and use the assigned type labels to measure whether the word type was changed\footnote{In case of multiple word type labels for a single token (e.g. noun and verb for \enquote{escape}), we only interpreted the perturbation as a word type change if the sets of word types of the original word and the new word were disjoint.} or not.

\begin{figure}%[htb]
    \centering
    \includegraphics[width=1.0\linewidth]{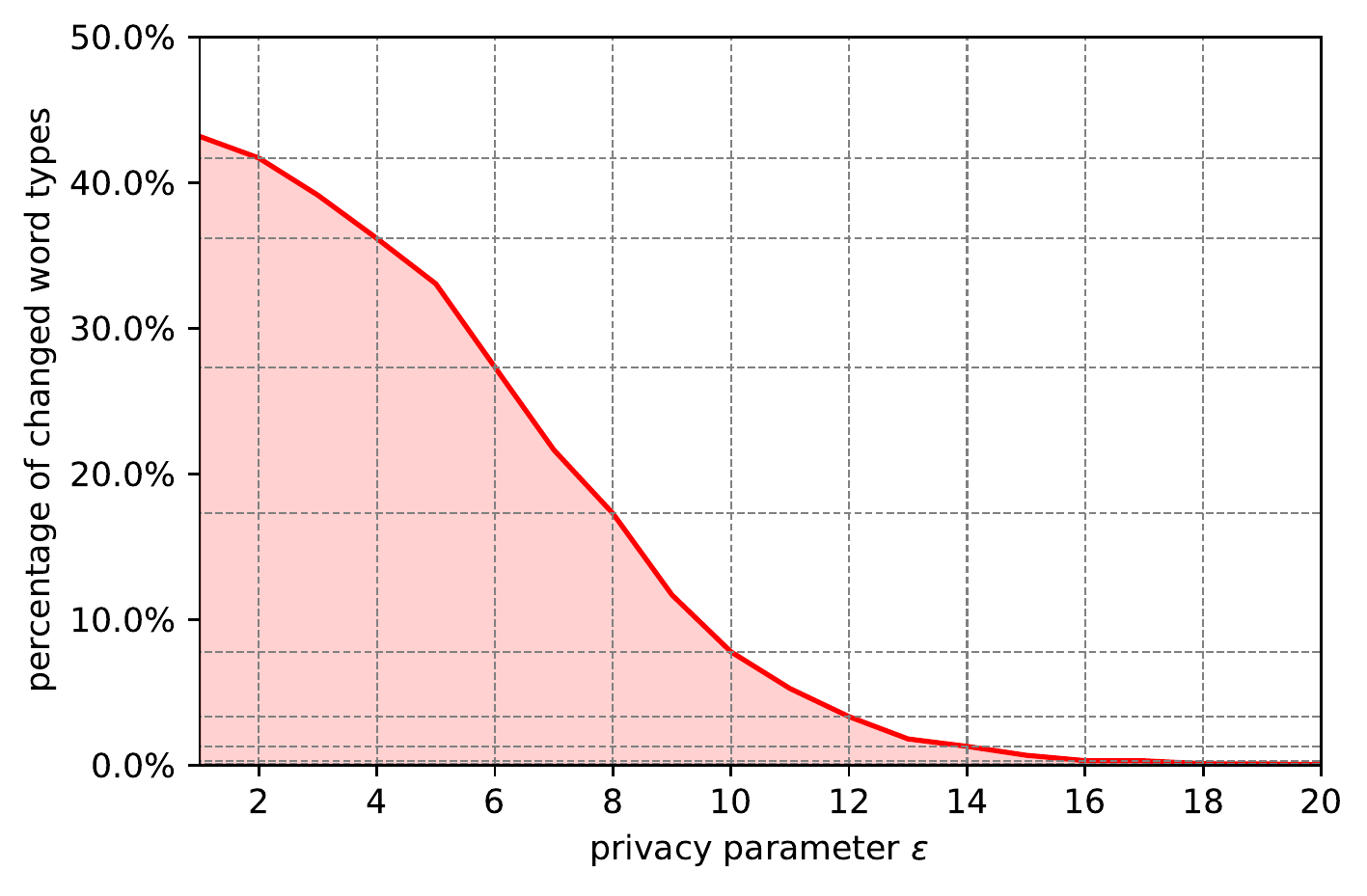}
    \caption{Percentage of word type changes caused by the mechanism introduced by \citet{feyisetan2020privacy}}
    \label{fig:wordtype}
\end{figure}

As \cref{fig:wordtype} shows, a significant percentage of word type changes occur even when using comparably large \(\epsilon\) values such as 8 or 10 that grant only little privacy protection according to our evaluation in \cref{sec:eval}: With 17.3\% and ~7.8\% of word types being changed with the respective epsilon values, a word type change and thus most likely a grammatical error would be induced at every 5.8th and 12.8th token, respectively. 

\paragraph{Lack of syntactic changes}
As described in \cref{sec:dpconstraints}, operating on a word-by-word basis causes severe limitations to the format of the perturbed output sentences.
Due to the imposed inflexibility of the text generation process, the discussed mechanisms lack the ability to rewrite given sentences by changing their syntactic properties such as word positioning and sentence length and thus mostly have to rely on lexical changes for obfuscating author-revealing features, which is highly unfavorable. For instance, if a person's writing style is characterized by heavy use of subordinate clauses resulting in very long sentences, it may be more effective to shorten sentences than merely changing individual words.

% According to research on authorship attribution,
Due to these limitations, word level methods may never achieve proper anonymization,
as even syntactic features alone without any semantic information are sufficient for authorship identification:
Notably, \citet{tschuggnall-specht-2014-enhancing} show that, given a collection of syntactic trees of texts written by various authors, individual style profiles can be learned to infer the authors of unseen sentences. Moreover, learned representations of syntax trees have proven to be effective for various authorship attribution tasks \citep{hitschler-etal-2017-authorship, zhang-etal-2018-syntax, syntactic_recurrent_authorship}. Consequently, an effective anonymization mechanism should be able to change the syntactic properties of its input texts in order to take away important clues that adversaries could exploit to identify authors.

\section{Anonymization through paraphrasing}
\label{sec:paraphrasing}
While existing works on text anonymization that focus on word level perturbations represent very important contributions, they have some significant weaknesses as described in \cref{sec:limitations}.
In the following, we attempt to address the identified problems
by proposing fine-tuning of large language models for paraphrasing
as an alternative text anonymization method.

\subsection{Generating paraphrases}
Authorship obfuscation has been framed as a paraphrasing problem in various works with different attempts to generate adequate rewritings \citep{rao-author, DBLP:conf/clef/KeswaniT0M16,bevendorff-etal-2019-heuristic, mahmood2019mutantx}.
While computational approaches do not always show satisfying results, \citet{mishari-crowdsourcing} demonstrate that rewritings of reviews gathered through crowdsourcing reflect strongly different stylometric features
from the source reviews while preserving the original content and concealing the author successfully.

Crowdsourcing is highly laborious and cannot always be applied in real-world scenarios.
Therefore, we aim at imitating the rewriting behavior of humans through a large-scale transformer-based \citep{NIPS2017_3f5ee243} language model: We fine-tune GPT-2 \citep{radford2019language} to generate paraphrases following the training procedure introduced by \citet{witteveen-andrews-2019-paraphrasing}.
The Stanford Natural Language Inference (SNLI) Corpus \citep{bowman-etal-2015-large}
provides training data consisting of pairs of sentences with five crowdsourced labels,
each indicating whether the two sentences are semantically entailed or not.
We construct a paraphrase dataset by only keeping sentence pairs with all five labels indicating entailment.

\subsection{Balancing privacy and utility}
As pointed out by \citet{brennan-adversarial}, the black box nature of authorship obfuscation via round-trip and consequently also monolingual translation
%\todo{used in our approach?} 
affects controllability of our system negatively.
Therefore, in the following we demonstrate how varying the temperature in the word sampling stage of GPT-2
can be used to inject noise into our model, hereby balancing the privacy-utility trade off.

% I changed the variable for words to x from w to maintain the notation from previous sections
In an autoregressive generative model, an output text $\vx = \parens{x_1,\ldots,x_n} $ is generated
by sampling the next word $x_i$ from conditional probabilities $\vv p_i=p(x_i \mid x_1,\ldots,x_{i-1}, \vz)$
modeled by the decoder network, where $\vz$ is context information
(e.g., representing an encoding of the input sentence to be obfuscated)
to initialize the decoder.
The vector $\vv p_i = (p_{i,j})_{j=1}^{\abs\V}$ represents the probabilities of producing
the $j$-th word $v_j$ of the predefined vocabulary $\V$ at the $i$-th position in the sequence.
The probabilities are typically obtained through the softmax function from a logit vector
$\vv u_i\in\Reals^{\abs\V}$ in the last layer of the decoder,
which can be controlled by a temperature parameter $T>0$ as follows:
\begin{equation}
  \label{eq:softmax}
  p_{ij} := \softmax\parens{\vv u_i}
    = \frac{\exp\parens*{\frac{u_{i,j}}{T}}}{\sum_k \exp\parens*{\frac{u_{i,k}}{T}}}
\end{equation}

% When generating text, the softmax function used to predict the probability of a token \(w_i\)
% out of the vocabulary \(\mathcal{V}\) given their unnormalized logit vector
% can be equipped with the temperature \(T\):
% \[
%   p_t(w_i) = \frac{\exp(\frac{w_i}{T})}{\sum_{j=1}^{|\mathcal{V}|} \exp(\frac{w_j}{T})}
% \]
A higher temperature $T$ results in a smoother distribution that brings the resulting probabilities of all words closer together and thus impacts the variability and probabilities of the resulting sentences.
In our experiments in \cref{sec:eval}, we vary the temperature when sampling text to evaluate this effect.

\paragraph{Sampling from $\softmax$ as \diffpriv mechanism}
Note that sampling from the softmax distribution with temperature $T$ 
can be interpreted as a \dlp mechanism, namely as an instance
of the \emph{Exponential mechanism} by \citet{mcsherry2007mechanism}.
It applies to both numerical and categorical data and
% \paragraph{The Exponential Mechanism}
% A very important and versatile \diffpriv mechanism
% that applies to both numerical and categorical data 
% is the Exponential mechanism by \citet{mcsherry2007mechanism}.
%For a given input, it randomly yields an output value based on a \enquote{rating} or \enquote{quality}
%that specifies the suitability of each possible output for the given input:
%While we could define sensitivity for the Laplace mechanism as distance on $\Reals^k$,
requires a \enquote{measure of suitability} for each possible pair of input and output values:
\begin{definition}[Quality function]
	\label{def:qualityfunc}
  A map $q:\X \times \Y \to \Reals$ is called \emph{quality function from $\X$ to $\Y$}
  where we interpret the value $q(x,y)$ as measure of suitability
  of an output $y\in\Y$ for a given input $x\in\X$.
	The \emph{sensitivity} $\Delta_q$ of the quality function $q$
  is its largest possible difference given two adjacent inputs,
  over all possible output values:
	\[
    \Delta_q := \max_{y \in \Y}\ \max_{x_1 \adj x_2} \big( q(x_1,y)-q(x_2,y) \big)
  \]
\end{definition}
Given an admissible rating function $q$ with finite sensitivity $\Delta_q$,
the Exponential mechanism is defined as follows:
%so the resulting Exponential mechanism will fulfill $(2\epsilon)$\-/\dlp.
%For a given input, the Exponential mechanism defines a probability distribution
%on the output space where the probability of an output for a given input
%is proportional to their exponentiated rating,
%multiplied by a privacy parameter $\epsilon$:
\begin{definition}[Exponential mechanism]
	\label{def:expmech}
	Let $\epsilon>0$ be a privacy parameter,
  and let $q:\X \times \Y \to \Reals$ be a rating function.
  The \emph{Exponential mechanism} is a random mechanism $\mathcal{E} : \X \to \Y$
  that is defined by the probability distribution function
	\begin{equation*}
    \label{eq:expmech}
		\pr{ \mathcal{E}(x) = y } =
    \frac{\exp\parens*{\frac\eps{2\Delta_q} q(x,y)}}
         {\int_{y'} \exp\parens*{\frac\eps{2\Delta_q} q(x,y')} \mathrm{d}y'}.
  \end{equation*}
	A discrete version of the Exponential mechanism for countable $\Y$
  can be obtained by replacing the integral with a sum; % from \cref{eq:expmech}
  it is thus defined by the probability mass function
	\begin{equation}
    \label{eq:expmech-discrete}
		\pr{ \mathcal{E}(x) = y } =
    \frac{\exp\parens*{\frac\eps{2\Delta_q} q(x,y)}}
         {\sum_{y'} \exp\parens*{\frac\eps{2\Delta_q} q(x,y')}}.
  \end{equation}
\end{definition}
The Exponential mechanism $\mathcal{E}$ fulfills $\epsilon$-\dlp
as shown by \citet[Theorem~6]{mcsherry2007mechanism}.

By comparing \cref{eq:softmax,eq:expmech-discrete},
we immediately recognize that sampling from the $\softmax$ probabilities
$\vv p_i=(p_{i,1},\ldots,p_{i,\abs\V})$
amounts to running an instance of the Exponential mechanism
with $\eps=2\Delta_q/T$ and the quality function
determined by the logits vector $\vv u_i\in\Reals^{\abs\V}$ as
\[
  q_i\parens*{(x_1,\ldots,x_{i-1},\vz), v_j} = u_{i,j},\quad 1 \leq j \leq \abs\V,
\]
at each iteration $i$ when sampling the next word $x_i$.
% with privacy parameter $\epsilon=1/T$
% and quality function $q_i\parens*{(x_1,\ldots,x_{i-1},\vz), v_j} = u_{i,j}$
% at each position $i$ in the output sequence. 
Therefore, our generative paraphrasing model naturally forms a locally \dyp mechanism
that also enjoys formal privacy guarantees:
The total privacy budget amounts to $\eps \cdot n$ where
% the sensitivity of the quality function is $\Delta\leq1$ and
$n$ is the length of the generated paraphrase.
%\todo{\textcolor{blue}{maybe mention overall privacy budget and how to bound it by limiting the output length of the decoder}}
Finally, note that we obtain a finite sensitivity $\Delta_q \leq 1$
by constraining the decoder layer so that the logits in its output
fulfill $0\leq u_{i,j} \leq 1$.

While this approach is still subject to the implications of the local model,
and its total privacy budget $\epsilon\cdot n$ may still grow linearly
in the length of the produced output,
it avoids the language and fixed output length constraints
of previous word level privacy mechanisms stated in \cref{sec:limitations}.
% \scriptsize \textcolor{blue}{No language constraints and no fixed-length output are indeed the case, but the growing privacy budget with longer sentences (right?) and general criticism of local model is also valid criticism for paraphrasing, maybe we should specify here}\normalsize

\section{Evaluation}
\label{sec:eval}
 
We argue that despite formal guarantees, the privacy preservation capabilities of mechanisms
that are deployed in real world applications should also be tested from a practical standpoint.
Previous works measure anonymization capabilities using a variety of evaluation metrics:
\citet{feyisetan2020privacy} use the privacy auditor proposed by \citet{10.1145/3292500.3330885},
%\todo{put citations directly next to their contribution (\ldots approaches such as (Feyisetan) use auditor by (Song+Shmatikov)\ldots), otherwise it looks like the auditor was \emph{proposed} by (Feyisetan)},
whereas \citet{xu2021utilitarian} measure the ability of an adversary to reconstruct the original sentence,
and \citet{DBLP:journals/corr/abs-2010-11947, xu2020differentially} count the amount of changed words.

Unfortunately, these methods are rarely tested under the scenario of a strong attacker
aiming to identify the authors of the obfuscated texts.
While \citet{feyisetan2019leveraging} measure the identification performance of an authorship attribution model,
their adversary \citep{articlekoppel} only relies on counting character 4-grams
and does not adequately reflect the capabilities of a strong attacker
who can train more powerful classifiers.
Besides, attacks are almost always evaluated only in a static (non-adaptive) setting,
meaning that the attack model is only trained on the original data and cannot adapt to the perturbed data.
Since any serious method should avoid \enquote{security (or privacy) by obscurity},
we must assume that the obfuscation mechanism is known to the attacker
who can easily create perturbed data themselves.
%\todo{also, attacks are almost always evaluated only in a non-adaptive setting}

% To address this, we evaluate  the privacy-utility trade off for 
In the following evaluation, we consider two exemplary methods following the word level framework,
namely perturbing Euclidean GloVe embeddings \citep{pennington-etal-2014-glove}
through Laplace noise as proposed by \citet{feyisetan2020privacy},
the perturbation of hierarchical Poincaré embeddings \citep{NIPS2017_59dfa2df}
through hyperbolic noise as proposed by \citep{feyisetan2019leveraging},
as well as our paraphrasing approach proposed in \cref{sec:paraphrasing}.
To address the discussed issues in previous evaluation methodologies,
we employ recent state-of-the-art methods to compare the privacy-utility trade-offs
and analyze the performance of the approaches not only in a static, but also in an adaptive setting.
% To address the evaluation-related issues mentioned in the previous paragraph,
% we specifically rely on recent systems achieving results that are state-of-the-art
% or very close to such performance as confirmed by their corresponding papers.
%\todo{Address how our evaluation is better regarding the issues raised in the preceding paragraphs.}

\begin{table*}[tb]
  \centering
  \small
  \caption{Performance of authorship and sentiment classifiers trained and evaluated on data generated by anonymization mechanisms as measured by MCC scores. \textbf{Best trade-offs} are identified by the \emph{relative gain} metric introduced in section \cref{sec:evalmetrics}}
  \label{tab:results:improper-dp}
  \begin{tabular}{ l c c c c c c c c c c c c c}
    \toprule
    & \multicolumn{1}{c}{original} & \multicolumn{4}{c}{GloVe embeddings} & \multicolumn{4}{c}{Poincaré embeddings} & \multicolumn{4}{c}{Paraphrase ($\epsilon=1/T$)}\\
    \cmidrule(lr){2-2} \cmidrule(lr){3-6} \cmidrule(lr){7-10} \cmidrule(lr){11-14}
    Privacy budget $\epsilon$ & \multicolumn{1}{c}{$\infty$} & 6 & 8 & 10 & 12 & 0.5 & 1 & 2 & 8 & 0.05 & 0.1 & 1.0 & 10\\
    \midrule
    \textbf{IMDb:} &\multicolumn{1}{c}{}&&&&&&&\\
    Author (static) & \multicolumn{1}{c}{0.98} & 0.12 & 0.20 & 0.28 & \textbf{0.33} & 0.87 & 0.88 & 0.88 & 0.89 &      0.19 & 0.21 & 0.22 & \textbf{0.22}\\
    Author (adapt.) & \multicolumn{1}{c}{0.98}  & 0.58 & 0.79 & 0.90 & 0.95 & 0.97 & 0.97 & 0.98 & 0.98 &   0.62 & \textbf{0.63} & 0.64 & 0.66\\ 
    Sentim. (static) & \multicolumn{1}{c}{0.71} & 0.21 & 0.32 & 0.43 & \textbf{0.50} & 0.53 & 0.52 & 0.52 & 0.53    & 0.37 & 0.40 & 0.40 & \textbf{0.42}\\ 
    Sentim. (adapt.) & \multicolumn{1}{c}{0.71} & 0.22 & 0.37 & 0.52 & 0.60 & 0.56 & 0.54 & 0.56 & 0.56    & 0.40 & \textbf{0.42} & 0.41 & 0.43\\  
    SBERT CS & \multicolumn{1}{c}{1.00} & 0.30 & 0.49 & 0.70 & 0.85 & 0.66 & 0.67 & 0.68 & 0.68    & 0.58 & 0.61 & 0.62 & 0.63\\
    PPL & \multicolumn{1}{c}{44.5} & 5003 & 3544 & 1414 & 512 & 431 & 384 & 330 & 310    & 37.2 & 34.8 & 34.4 & 33.9\\
    \midrule
    % \addlinespace
    \textbf{Yelp:} &\multicolumn{1}{c}{}&&&&&&&\\
    Author (static) & \multicolumn{1}{c}{0.80} & 0.12 & 0.23 & \textbf{0.40} & 0.49 & 0.59 & 0.61 & 0.60 & 0.62   & 0.22 & 0.35 & 0.37 & 0.38\\
    Author (adapt.) & \multicolumn{1}{c}{0.80} & 0.32 & 0.47 & 0.62 & 0.68 & 0.72 & 0.73 & 0.73 & 0.75   & \textbf{0.35} & 0.35 & 0.37 & 0.39\\ 
    Sentim. (static) & \multicolumn{1}{c}{0.51} & 0.14 & 0.20 & \textbf{0.27} & 0.33 & 0.35 & 0.37 & 0.36 & 0.37    & 0.20 & 0.21 & 0.23  & 0.24\\ 
    Sentim. (adapt.) & \multicolumn{1}{c}{0.51} & 0.17 & 0.26 & 0.34 & 0.43 & 0.44 & 0.45 & 0.45 & 0.46    & \textbf{0.32} & 0.30 & 0.30 & 0.33\\  
    SBERT CS & \multicolumn{1}{c}{1.00} & 0.29 & 0.43 & 0.60 & 0.76 & 0.35 & 0.37 & 0.38 & 0.38 &   0.49 & 0.51 & 0.54 & 0.54\\
    PPL & \multicolumn{1}{c}{99.7} & 13427 & 8555 & 3061 & 1534 & 1248 & 1232 & 1155 & 1116 & 148 & 143 & 138 & 132\\
    
    \bottomrule
  \end{tabular}
\end{table*}

\subsection{Evaluation metrics}
\label{sec:evalmetrics}
We argue that an anonymization mechanism deployed in real world applications should provide protection against advanced deanonymization attacks, preserve the core information of the original data (e.g., sentiment for product reviews), be semantically similar to the original sentences and of high quality in terms of language.

We measure the first two properties using both static (i.e., trained on source data) and adaptive (i.e., trained on data perturbed by the respective mechanism) BERT-based \citep{devlin-etal-2019-bert} author and sentiment classifiers by fine-tuning the pretrained language model's top three layers and using a two-layer classifier for the author and sentiment labels.
\todo{Is it the same (shared) two-layer classifier for author and sentiment, or two separate classifiers?}
BERT has proven to be successful for both sentiment classification \citep{bertclassification} and authorship attribution \citep{Fabien2020BertAAB} and thus represents a suitable model for both tasks.
We report all classification results in terms of Matthews Correlation Coefficient (MCC) \citep{matthews_comparison_1975,gorodkin_comparing_2004}.
An MCC score of +1 means perfect predictions whereas 0 indicates random guessing.
MCC is more suitable to assess classification performance than accuracy \citep{chicco2020advantages} 
as it is not easily fooled by biased classifiers in case of imbalanced datasets.

To assess the trade-off between attack (authorship attribution) and utility (sentiment analysis),
we measure each method's \emph{relative gain} based on the original and obfuscated classification scores:
Let $A_o, S_o$ represent the MCC scores of the author and sentiment classifiers
based on the original data, and similarly, let $A_p, S_p$ represent the scores on perturbed data
normalized to the range $[0,1]$. Then we define its relative gain as $\gamma:=S_p/S_o - A_p/A_o$.
% That way, a \enquote{perfect} trade-off indicating no author information but full utility preservation
% achieves $\gamma=1$, whereas full author information but no utility preservation results in $\gamma=-1$.
% Methods that do not change the data get $\gamma=0$.

To measure semantic similarity between the anonymized and original sentences, we compute the cosine similarity of their representations obtained by SBERT \citep{reimers-gurevych-2019-sentence}, which is a model that has been optimized for capturing semantic similarity between textual inputs.
For language quality, we compute the average perplexity (PPL) of the pretrained GPT-2 \citep{radford2019language} over the output sentences of each model.

\subsection{Implementation Details}

We implement both mechanisms proposed in the papers by \citet{feyisetan2020privacy,feyisetan2019leveraging}
using numpy.
Concretely, we use 50-dimensional GloVe \citep{pennington-etal-2014-glove} vectors as our Euclidean embeddings and train 50-dimensional Poincaré embeddings on our own.
For the latter, we extract \(\sim\)1,300,000 word tuples representing hypernymy relationships for IMDb and \(\sim\)1,800,000 tuples for Yelp from WebIsADB\footnote{Terms of use and license information: \cref{sec:dataterms}} \citep{seitner-etal-2016-large} by removing words with less than 10 occurrences and keeping only tuples contained in the GloVe vocabulary as well as the respective review dataset\footnote{As the procedure was not fully described in the paper, we increased (by factor \(\geq\) 10) the training data of the original work, hereby having a larger vocabulary and more variation in the perturbations. We do so to minimize the risk of bad results merely due to implementation issues.}.
%\todo{rephrase? (this sentence is a bit long, same for footnote)}

When encountering out-of-vocabulary (OOV) words, \cref{alg:wordleveldp} cannot assign embeddings and thus not perturb them, which violates \dlp. Also, merely removing the words does not change this as it changes the length of the output text while \dlp is only fulfilled for texts of the same length. For GloVe embeddings, a relevant effect in terms of experimental results is not present as the large vocabulary covers almost all words we encounter. The vocabulary size of our Poincaré embeddings is however limited (\(\sim\) 10,000) and, following \citet{feyisetan2019leveraging}, does not contain stopwords. As we aim to compare methods outputting human-readable texts and the removal of stopwords clearly affects readability, we instruct the mechanism to simply ignore OOV words. The results for removing OOV words can be found in \cref{tab:results:proper-dp} in the appendix. 

For GPT-2 and BERT, we use the pretrained checkpoints from the HuggingFace transformers library (\emph{gpt2, bert-base-uncased}; 117M, 110M parameters) and fine-tune each instance on a single NVIDIA T4 GPU.
%To evaluate the approaches fairly, we deal with this problem in two ways: Our first approach with results reported in table \ref{tab:results:proper-dp} removes out-of-vocabulary words from the output texts whereas our second approach reported in table \ref{tab:results:improper-dp} keeps them without perturbation.

\subsection{Datasets}
We conduct experiments using IMDb movie reviews \citep{maas-etal-2011-learning} and Yelp business reviews\footnote{\url{https://www.yelp.com/dataset/}} which contain author and sentiment labels in the form of ratings on the scale of 1-10 and 1-5, respectively. For both sources, we keep data from ten users with the most reviews, hereby obtaining dataset sizes of 10,000 for IMDb and 15,729 for Yelp. We simplify sentiment labels by rating movie reviews with \(\geq 5\) points and business reviews with \(\geq 3\) points as positive and the rest negative.
%\todo{check: positive $\leftrightarrow$ negative?}

\subsection{Results}
%\todo{Redundant? (duplicated from "Implementation" section)}

\cref{tab:results:improper-dp} shows that paraphrasing significantly outperforms word-level mechanisms in terms of protection against adaptive adversaries. When evaluating privacy and utility for static classifiers, it becomes apparent that small perturbations are enough to trick author classifiers. Therefore, for static classifiers, mechanisms with weak word-level perturbations caused by smaller \(\epsilon\) values show an equal trade-off on IMDb and a slightly better trade-off on Yelp reviews as they better preserve the sentiment than the stronger changes caused by our model. Notably, paraphrasing shows better semantic preservation as well as higher language quality as measured by PPL when comparing it to word-level mechanisms calibrated for comparable privacy protection against the author classifier. This is also visible in the exemplary outputs provided in \cref{tab:sentences}.

\section{Related work}
\label{sec:rel-work}

\paragraph{Other \dlp mechanisms for text}
Earlier mechanisms for \dyp text obfuscation settled for simpler output representations: % that are not human-readable:
\citet{weggenmann2018syntf,fernandes2019generalised} employ \bow models and produce term-frequency vectors as output.
Similarly, obfuscated dense vector representations are obtained in \cite{beigi2019i}
by perturbing the output of an encoder network.
While not human-readable, these vector representations can be shared for automated processing,
such as topic or sentiment inference and machine learning.
%\todo{Mention Bo et al. + differences? Yes :)} 
To generate human-readable text, \citet{bo-etal-2021-er} employ an encoder-decoder model similar to ours, but without paraphrasing,
and sample output words using (a two-set variant of) the Exponential mechanism
\cite{mcsherry2007mechanism}. \citet{dp-vae-weggenmann} propose a differentially private variation of the variational autoencoder and use it as a sequence-to-sequence architecture for text anonymization.

\paragraph{Authorship obfuscation without \dlp}
Approaches not following \dlp range from rule-based algorithms relying on human-engineered text perturbations such as synonym replacements or word removals \citep{bevendorff-etal-2019-heuristic, mahmood2019mutantx} to methods incorporating deep learning. Models of the latter typically incorporate discriminator networks to penalize generating author-revealing information \citep{shetty2018a4nt, xu2019privacy}. Similar to \dlp mechanisms, previous work is concerned with learning private text vector representations \citep{coavoux-etal-2018-privacy}.

\paragraph{Differentially private optimization}

Differentially private optimization algorithms such as DP-SGD and related methods \citep{dpsgd-song, dpsgd-bassily, dpsgd} have emerged as effective methods for protecting the training data of a model. Recent work has shown that both generative and discriminative language models can effectively be trained with these optimization approaches \citep{li2021large, lm-dp-finetuning} and therefore represent an important contribution for protecting against data leakage of language models \citep{lm-dataleak, lm-extractdata}. These methods can be seen as complementary to the approaches discussed in this paper which protect data during inference.

\section{Conclusion}
We discussed and demonstrated the weaknesses of word level \dlp mechanisms and proposed a paraphrasing model circumventing most of these. We find that our approach outperforms word level mechanisms in terms of protection against adaptive adversaries, while the latter should be favored against weaker adversaries. Future work could address integrating auxiliary adversarial losses to paraphrasing systems or enabling paraphrases that better preserve the core information of the source text.
\label{sec:conclusion}

\section{Ethical Considerations}

\paragraph{Abuse of Anonymization Mechanisms}
Text Anonymization is an important field of research for the protection of privacy of individuals as well as for enabling freedom of speech. Still, anonymization mechanisms may be exploited for negative causes. Specifically, guaranteed anonymity on the internet might lead individuals to spread hate speech. Furthermore, mechanisms as ours can be used to anonymously generate and spread fake reviews or fake news. Important areas of research fighting these problems include hate speech and toxicity detection \citep{djuric2015hate, macavaney2019hate} as well as fake review detection \citep{mukherjee2013fake, barbado2019framework} and fake news detection \citep{shu2017fake, ruchansky-csi}.

\paragraph{Bias in Large Language Models} Large language models such as GPT-2, which our proposed approach is based on, often inherit biases towards various demographics from the large amount of data they are trained on \citep{sheng-etal-2019-woman, abid2021large}. These biases can cause unforeseen effects when generating language output and could potentially alter statements of authors whose texts are being anonymized. An increasing amount of work is aiming to understand and tackle such biases in language models \citep{NEURIPS2020_92650b2e, liang2021towards}.

\paragraph{Evaluation Fairness}
In this paper, we evaluate our approach experimentally and compare its performance to mechanisms proposed in previous research works. Since no code was publicly released for the approaches we are comparing ours to, we implemented the mechanisms ourselves. While we replicated the original systems as close as possible to the description in the papers using all the information available, we cannot guarantee that they are exactly the same as not all the information about preprocessing and implementation details is publicly available.

\section*{Acknowledgements}

We thank Zhijing Jin for the helpful discussions about the presentation of our results and the design of our paper.
\bibliography{anthology,custom}
\bibliographystyle{acl_natbib}

\appendix

\section{Appendix}
\label{sec:appendix}

\subsection{Information about terms of use for data}
\label{sec:dataterms}

In this section, we provide information and references about the terms of use and licenses of each dataset we are using.

\paragraph{WordNet}
Wordnet can be downloaded and accessed online without specifically requesting access and can be used for research and also commercial applications in accordance with the WordNet 3.0 license: \url{https://wordnet.princeton.edu/license-and-commercial-use}

\paragraph{WebIsADB}
WebIsADB can be downloaded and accessed online without specifically requesting access. The dataset is licensed under a Creative Commons Attribution-Non Commercial-Share Alike 3.0 License:  \url{http://creativecommons.org/licenses/by-nc-sa/3.0/}

\paragraph{IMDb}
IMDb movie reviews can be downloaded and accessed online without specifically requesting access. Unfortunately, we could not find information about license specifications. More information is available at \url{https://ai.stanford.edu/~amaas/data/sentiment/}

\paragraph{Yelp}
Researchers aiming to use the Yelp dataset have to sign the terms of use (\small\url{https://s3-media3.fl.yelpcdn.com/assets/srv0/engineering_pages/bea5c1e92bf3/assets/vendor/yelp-dataset-agreement.pdf}\normalsize). For commercial use, researchers should contact Yelp via \url{dataset@yelp.com}. More information is available at \url{https://www.yelp.com/dataset}.

\begin{table*}
\centering
\caption{Exemplary output of anonymization mechanisms for Yelp data}
    \label{tab:sentences}

    \begin{tabular}{l}
        
    \hline
Exemplary Reviews for Yelp\\
\hline
\\
         \textbf{Original:}\\
         This store is so adorable . In addition to baked goods they offer sandwiches for breakfast and lunch .\\ The turkey sandwich was excellent . The textures were perfect though, especially for\\ the almond amaretto cookie . It had the right balance of chewy\\ with a slight amount of crunch. \\
         
         \\
         \textbf{Euclidean embedding perturbations (\(\epsilon = 8\)):}\\
         designated store is work adorable making top tubular continue watered goods do offer salad ranging\\ breakfast filling 5,000-a carries original turkey sandwich was excellent parts national textures\\ were play never neighbors with for part mustard amaretto cookie hatred\\ make had a direction balance end sugary another a erratic amounts of one-off today \\
         \\
         \textbf{Euclidean embedding perturbations (\(\epsilon = 10\)):}\\
         fact store is 're granny his in health they dish goods kept offer sandwiches giving dinner besides\\ lunch result the turkey sandwich given delivering . the textures ten captures .\\ then especially own the apricot izola cookie on be had the right footing of chewy\\ with a slight amount in crunch at \\
         \\
         \textbf{Poincaré embedding perturbations (\(\epsilon = 1\)):}\\
         this flag is so adorable . in abundance to waffles chunk they many slimy for eggs and \\ peppery . the vindaloo stickers was excellent . the blt were splurge gun ,\\ especially for the quail amaretto crunch . it had the quantity\\ observation of crisp with a trotter simple of crunch . \\
         \\
         \textbf{Poincaré embedding perturbations (\(\epsilon = 2\)):}\\
         this patient is so adorable . in stuff to asparagus walk-up they con jets for pricy and tamale .\\ the petite cook was excellent . the rang were steal train , especially for the updated amaretto soak .\\ it had the say many of containing with a slight land of cans .  \\
         \\
         \textbf{Paraphrased (\(\epsilon = 0.1\)):}\\
         There is a cute store. There is a sandwich being served by the sandwich shop. The sandwich is tasty.\\ The two textures are alike. There were chews on the chem.\\
                  \\
         \textbf{Paraphrased (\(\epsilon = 1\)):}\\

         There's adorable store in this photo. In addition to baked goods they offer sandwiches for\\ breakfast and lunch. This was a great sandwich. \\The desserts taste delicious! The food was chewy. \\
         \\
         \hline

    \end{tabular}

\end{table*}

\begin{table*}[tb]
  \centering
  \small
  \caption{Results for hyperbolic perturbations \cite{feyisetan2019leveraging} when removing out-of-vocabulary words% with GloVe and Poincaré embeddings \cite{feyisetan2020privacy,feyisetan2019leveraging}
              \todo{including stopwords in the perturbation} \todo{use similar (better: same) \eps values?}.}
  \label{tab:results:proper-dp}
  \begin{tabular}{ l  c c c c c }
    \toprule
    & \multicolumn{1}{c}{original} &  \multicolumn{4}{c}{Poincaré embeddings} \\
    \cmidrule(lr){2-2} \cmidrule(lr){3-6}
    Privacy budget $\epsilon$ & \multicolumn{1}{c}{$\infty$} & 0.5 & 1 & 2 & 8  \\
    \midrule
    \textbf{IMDb:} &\multicolumn{1}{c}{}&&&\\
    Author MCC (static) & \multicolumn{1}{c}{0.98} & 0.03 & 0.12 & 0.07 & 0.12  \\
    Author MCC (adapt.) & \multicolumn{1}{c}{0.98}  & 0.67 & 0.69 & 0.68 & 0.69    \\ 
    Sentim.~MCC (static) & \multicolumn{1}{c}{0.71} & 0.27 & 0.30 & 0.31 & 0.28    \\
    Sentim.~MCC (adapt.) & \multicolumn{1}{c}{0.71} & 0.35 & 0.40 & 0.38 & 0.39    \\
    SBERT CS & \multicolumn{1}{c}{1.00} & 0.32 & 0.33 & 0.33 & 0.34 \\
    \midrule
    % \addlinespace
    \textbf{Yelp:} &\multicolumn{1}{c}{}&&&\\
    % \addlinespace[.5ex] % add some space for extra structure
    % changed row label to align (easier to read)
    Author MCC (static) & \multicolumn{1}{c}{0.80}  & 0.14 & 0.14 & 0.15 & 0.16 \\
    Author MCC (adapt.) & \multicolumn{1}{c}{0.80} & 0.32 & 0.35 & 0.34 & 0.36  \\
    Sentim.~MCC (static) & \multicolumn{1}{c}{0.51} & 0.17 & 0.18 & 0.20 & 0.20  \\
    Sentim.~MCC (adapt.) & \multicolumn{1}{c}{0.51} & 0.21 & 0.23 & 0.24 & 0.25  \\  
    % \addlinespace[.5ex]
    SBERT CS & \multicolumn{1}{c}{1.00} & 0.54 & 0.54 & 0.56 & 0.57 \\
    \bottomrule
  \end{tabular}
\end{table*}

\end{document}